\documentclass[submission, Phys]{SciPost}

\hypersetup{
	colorlinks,
	linkcolor={red!50!black},
	citecolor={blue!50!black},
	urlcolor={blue!80!black}
}

\usepackage[bitstream-charter]{mathdesign}
\DeclareSymbolFont{usualmathcal}{OMS}{cmsy}{m}{n}
\DeclareSymbolFontAlphabet{\mathcal}{usualmathcal}

\begin{document}
	
	\begin{center}{\Large \textbf{
				Universal non-Hermitian flow in one-dimensional PT-symmetric  
        quantum criticalities
	}}\end{center}
	
	\begin{center}
		Xin-Chi Zhou\textsuperscript{1,2} and
		Ke Wang\textsuperscript{3$\star$}

        ${}^\star$ {\small \sf kewang07@uchicago.edu}

	\end{center}
 
	\begin{center}
		{\bf 1} International Center for Quantum Materials, School of Physics, Peking University, Beijing 100871, China
		\\
  		{\bf 2} Hefei National Laboratory, Hefei 230088, China
            \\
		{\bf 3}  Kadanoff Center for Theoretical Physics,  James Franck Institute, Department of physics, University of Chicago, Chicago, IL 60637, USA

	\end{center}
	
	\begin{center}
		\today
	\end{center}
	
	
	\section*{Abstract}
	{\bf
         The critical point of a topological phase transition is described by a conformal field theory (CFT), where the finite-size corrections to the ground state energy are uniquely related to its central charge. We study the finite-size scaling of the energy of non-Hermitian Su-Schrieffer-Heeger (SSH) model with parity and time-reversal symmetry ($\mathcal{PT}$) symmetry. We find that under open boundary condition (OBC), the energy scaling $E(L)\sim c/L$  reveals a negative central charge $c=-2$ at the non-Hermitian critical point, indicative of a non-unitary CFT. Furthermore, we discover a universal scaling function capturing the flow of a system from Dirac CFT with $c=1$ to a non-unitary CFT with $c=-2$. The scaling function demonstrates distinct behaviors at topologically non-trivial and trivial sides of critical points. Notably, within the realm of topological criticality, the scaling function exhibits an universal rise-dip-rise pattern, manifesting a characteristic singularity inherent in the non-Hermitian topological critical points. The analytic expression of the scaling function has been derived and is in good agreement with the numerical results.
	}

	\vspace{10pt}
	\noindent\rule{\textwidth}{1pt}
	\tableofcontents\thispagestyle{fancy}
	\noindent\rule{\textwidth}{1pt}
	\vspace{10pt}
	
\section{Introductions}
Non-Hermitian systems~\cite{ashida2020,bergholtz2021}, which can effectively describe various physical phenomena such as quasiparticles in many-body systems~\cite{kozii2017,shen2018,papaj2019,nagai2020}, classical systems with gain and loss~\cite{el-ganainy2018a,zhou2018,ezawa2019,helbig2020,ghatak2020,hofmann2020}, and dissipative quantum gas of light~\cite{ozturk2021}, have attracted considerable interest in recent years. One of the striking features of non-Hermitian systems is the emergence of novel phenomena that have no Hermitian analogues, such as the non-Hermitian skin effect~\cite{yao2018,lee2019,lee2019a,song2019a,okuma2020,li2020,kawabata2020,zhang2022a,ren2022,zhao2023,kawabata2023,wang2023a,lee2024}, which violates the conventional bulk-edge correspondence of topological systems~\cite{kunst2018,xiong2018,jin2019,kunst2019,edvardsson2019,yokomizo2019,herviou2019,yang2020a,zhang2020,koch2020,kawabata2020a,xi2021,hu2021,fu2023,sayyad2023,okuma2023}. Another important aspect of non-Hermitian systems is the role of parity and time-reversal ($\mathcal{PT}$) symmetry~\cite{bender1998,bender2002,longhi2009,hu2011a,regensburger2012,feng2014,konotop2016,el-ganainy2018a,longhi2018}. A $\mathcal{PT}$-symmetric non-Hermitian Hamiltonian can have a completely real spectrum, even though it is not Hermitian. However, $\mathcal{PT}$-symmetry can be spontaneously broken when some eigenvalues become complex, leading to the formation of exceptional points where eigenstates and eigenvalues coalesce~\cite{heiss2012}. $\mathcal{PT}$-symmetry can also enrich the physics of Hermitian systems, as exemplified by non-unitary quantum dynamics~\cite{McDonald2018,Clerk2019,dora2020,gopalakrishnan2021,legal2023,morales-tejera2023}, $\mathcal{PT}$-symmetric quantum criticality~\cite{ashida2017,arean2020,kawabata2021,dora2022,grunwald2022,tu2022}, and non-unitary conformal field theories (CFTs)~\cite{npb84VBelavin,bianchini2014,chang2020,hsieh2023}.

Non-Hermitian lattice models can exhibit quantum criticality that is described by non-unitary CFTs, which are characterized by a negative central charge $c<0$. In the unitary representations of the Virasoro algebra, the absence of negative-norm states imposes constraints on the central charge and the highest conformal weight $h$\cite{fran12conformal}, resulting in a non-negative central charge. However, in the non-unitary representations, these constraints are relaxed, and negative central charges are possible. The central charge is a universal quantity that manifests in both the entanglement entropy~\cite{Calabrese_2009} and the conformal spectrum~\cite{prl86Cardy,KOO1994459,prl86Affleck}. Ref.~\cite{chang2020} used the scaling relation of the entanglement entropy $S$ with respect to the system size $L$, expressed as $S\sim c\log L$, to identify the negative central charge. Nevertheless, the scaling relation of the many-body ground-state energy $E(L)\sim c/L$, which also reflects the central charge, has not been verified for non-unitary CFTs. Moreover, for a generic lattice model whose Hermicity is controlled by a non-Hermitian parameter, varying this parameter induces a transition from a Hermitian to a non-Hermitian regime, and the CFT changes from unitary ($c>0$) to non-unitary ($c<0$). It is still unclear how the CFT evolves from unitary to non-unitary as a function of the non-Hermiticity parameter.

\begin{figure}
\centering{}\includegraphics[scale=1.2]{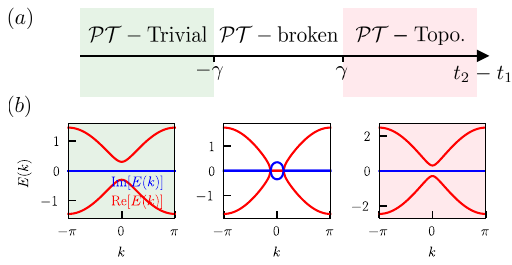}\caption{\label{Fig1}(a) The phase diagram of $\mathcal{PT}$ SSH model as a function of
$t_{2}-t_{1}$ and $\gamma$. (b) The dispersion of $\mathcal{PT}$-symmetric
topological phase, $\mathcal{PT}$-broken phase and $\mathcal{PT}$-symmetric
trivial phase, respectively.}
\end{figure}

In this work, we study the $\mathcal{PT}$-symmetric quantum criticality using non-Hermitian SSH model as a prototypical example. The activation of the non-Hermitian parameter leads to the bifurcation of the single gapless critical point between topological and trivial phase into two discrete critical points [see Fig.~\ref{Fig1}(a)]. One is the non-Hermitian topological critical point separating the $\mathcal{PT}$ symmetric topological phase from $\mathcal{PT}$ broken phase; the other is the non-Hermitian trivial critical point separating $\mathcal{PT}$ topologically trivial phase from $\mathcal{PT}$ broken phase.

We find that the scaling of ground state energy, expressed as $E(L)\sim c/L$, is still valid at the two critical points in the non-Hermitian regime and reveals the negative central charge $c=-2$ of non-hermitian lattice model in the open boundary. 

Moreover, we discover a universal scaling function that encapsulates the evolution of a system from unitary to non-unitary CFTs as the non-Hermitian scale varies. The scaling function is a consequence of both non-Hermicity and non-trivial topology of the system\cite{Kamenev,wang1,wang2,M1}. Specifically, turning on the non-Hermitian parameter, the system undergoes the transition from a Hermitian to a non-Hermitian one, leading the CFT to flow from unitary ($c>0$) to the non-unitary regime ($c<0$). Notably, the scaling function demonstrates distinct behaviors on the topologically non-trivial and trivial sides of the non-Hermitian critical points. Within the side of non-Hermitian topological phase transition, the scaling of flow exhibits an exotic rise-dip-rise pattern attributed to a {\it universal singularity}, reflecting its sensitivity to the topology of the quantum criticality. We further derive the analytic expression of the scaling function, elucidating a profound linkage between the non-Hermitian topological edge mode and the unique quantization condition embedded in the non-unitary CFT. The analytic findings also show that the scaling function maintains its universality across a broad spectrum of non-Hermitian systems with linear dispersion, regardless of microscopic details of the model.

\section{Model}

\subsection{$\mathcal{PT}$-symmetric SSH model}

We investigate the $\mathcal{PT}$-symmetric SSH model~\cite{liang2013,klett2017,lieu2018,chang2020}, which is described
by the Hamiltonian
\begin{equation}
H=\sum_{j}(t_{1}c_{A,j}^{\dagger}c_{B,j}+t_{2}c_{B,j}^{\dagger}c_{A,j+1}+\mathrm{h.c.})+i\gamma\sum_{j}(c_{A,j}^{\dagger}c_{A,j}-c_{B,j}^{\dagger}c_{B,j}),\label{Eq:PTSSH}
\end{equation}
where $c_{\sigma,j}^{\dagger}(c_{\sigma,j})$ creates (annihilates)
a particles at site $j$ on sublattice $\sigma=\{A,B\}$, $t_{1}$
and $t_{2}$ denote real hopping coefficients. The last term in eq.~\ref{Eq:PTSSH} is a purely imaginary staggered on-site potential which accounts for the gain and loss with strength $\gamma$. $\gamma$ is the origin of non-Hermicity of the system and tunes the strength of the non-Hermicity, when $\gamma=0$, the system reduces to the Hermitian SSH model. Under periodic boundary
conditions (PBC), then non-Hermitian Bloch Hamiltonian takes the form $H=\sum_{k}\Psi_{k}^{\dagger}h_{k}\Psi_{k}$,
where $\Psi_{k}^{\dagger}=[c_{A,k}^{\dagger},c_{B,k}^{\dagger}]$, with $c_{\sigma,k}^{\dagger}$ ($c_{\sigma,k}$) creating (annihilating) a particle with momentum $k$ on the sublattice $\sigma$, and the bloch Hamiltonian is written as $h_{k}=(t_{1}-t_{2}\cos k)\sigma_{x}+t_{2}\sin k\sigma_{y}+i\gamma\sigma_{z}$.
Here $\sigma_{i=x,y,z}$ are Pauli matrices. The dispersion relation
follows as
\begin{equation}
E(k)=\pm\sqrt{t_{1}^{2}+t_{2}^{2}-2t_{1}t_{2}\cos k-\gamma^{2}}.
\end{equation}
We briefly review the properties of $\mathcal{PT}$-symmetric SSH
model. For $\gamma\neq0$, the system exhibits $\mathcal{PT}$ symmetry
$\sigma_{x}h_{k}\sigma_{x}=h_{k}^{*}$. When $\gamma$ is small, the dispersion is purely real and when $\gamma$ is large enough, the spectrum becomes complex [Fig.~\ref{Fig1}(b)]. The model undergoes three
distinct phases as the $\gamma$ increases as illustrated in Fig.~\ref{Fig1}: (i) $\mathcal{PT}$-symmetric topological phase for
$t_{2}-t_{1}>\gamma$, (ii) $\mathcal{PT}$-symmetric trivial phase
for $t_{2}-t_{1}<-\gamma$ and (iii) $\mathcal{PT}$-broken phase
for $|t_{2}-t_{1}|<\gamma$. In the $\mathcal{PT}$-symmetric topological
phases, the bulk spectrum is fully gapped and completely real for the PBC. For the OBC, there are a pair of
edge states with pure imaginary energies $E=\pm i\gamma$ and the rest of the spectrum are also purely real. In the
$\mathcal{PT}$-broken phases, the spectrum under both OBC and PBC becomes complex and gapless.
In this work, we first investigate the non-Hermitian critical point between $\mathcal{PT}$-symmetric
topological phase and $\mathcal{PT}$-broken phase $t_{2}-t_{1}=\gamma$, where the spectrum is gapless. And then we investigate how the criticality evolves from zero $\gamma$ to finite $\gamma$ by gradually turning on the $\gamma$ while keeps $|t_2 - t_1|=\gamma$.

\section{Critical points and non-unitary CFTs}
The operator-state correspondence is an essential feature of CFTs. Any states in the CFT spectrum can be constructed from the local primary operators and their descendants. Furthermore, in the finite-size system, the energy of eigenstates can be expressed in terms of the scaling dimension of corresponding local operators. For example, the eigenvalues in open (periodic) boundary lattice models $E^{P(O)}_\alpha$ are given by\cite{zouyijian},
\begin{equation}
\label{corr}
E^O_\alpha(L)= A_O+ \frac{\pi v_F}{L} \left(h_\alpha- \frac{c}{24} \right)
 ,\quad E^P_\alpha(L)= A_P+ \frac{\pi v_F}{L} \left(\Delta_\alpha- \frac{c}{6} \right).
\end{equation} 
Here, $c$ is the central charge, $v_F$ is the Fermi velocity, and $\Delta_\alpha$ and $h_\alpha$ are scaling dimensions. The constants $A_{P/O}$ are non-universal {and depend on microscopic detail of the model}. Importantly, the universal relations expressed in Eq.~\ref{corr} enable us to compute the central charge of CFTs and the scaling dimensions of operators. {The model dependent non-universal constants $A_{P/O}$ can be explicitly attributed to contributions from the bulk dispersion and boundary excitation.} The finite-size scaling of the ground state energies in the open (periodic) boundary condition $E^{O(P)}(L)$ are given by 
\begin{equation}
\label{eq:critical-scaling}
E^{O}(L)= L \epsilon + b -\frac{\pi}{24}\frac{v_F c}{L} + \mathcal{O}(L^{-2}) ,\quad E^{P}(L)= L \epsilon -\frac{\pi}{6}\frac{v_F c}{L} + \mathcal{O}(L^{-2}),
\end{equation} 
where $\epsilon$ is the average bulk energy per particle and $b$ is the size-independent boundary term. For the PBC, there is no boundary excitation,resultin in no contribution from $b$.

In this work, we show that the finite size scaling Eq.~\ref{eq:critical-scaling} is also valid for non-Hermitian case with a negative central charge. Furthermore, we derive the universal scaling function that characterizes the quantum critical points from Hermitian to non-Hermitian, both numerically and analytically.  Specifically, we introduce a parameter $\Lambda$ that controls the non-Hermitian scale of the system, leading to the generalized finite-size scaling form:
\begin{equation}
\label{eq:universalFunc}
E^{O}(L,\Lambda)= L \epsilon + b -\frac{v_F}{L}f(\Lambda),
\end{equation} 
where the $f(\Lambda)$ describes the universal flow from the unitary CFT to the non-unitary CFT.

\subsection{Numerical method to obtain scaling function}
We first outline the numerical method to obtain the finite-size scaling
function in Eq.~\ref{eq:critical-scaling} and Eq.~\ref{eq:universalFunc}. Here the ground state energies $E^{O}(L)$ and $E^{P}(L)$ are the sum of the energies of all the occupied eigenstates, which can be obtained by numerically diagonalizing the Hamiltonian under open and periodic boundary condition, respectively. The bulk energy per particle, denoted as $\epsilon$, is obtained via integration across the entire Brillouin zone for all filled bands, which is calculated by
\begin{equation}
    \epsilon = -\frac{1}{2\pi} \int_{\mathrm{BZ}} dk \sum_s E_s(k),
\end{equation}
where $E_s(k)$ is the dispersion and $s$ is the index for the occupied band. The boundary term, which is a model-dependent constant, can be approximately obtained by 
\begin{equation}
    \Tilde{b}\approx E^{O}(L_0)-L_0\epsilon,
\end{equation}
with $L_0$ representing a sufficiently large system size, and $\Tilde{b}$ denotes the approximated boundary term. Subsequently, the finite-size scaling function can be evaluated through
\begin{equation}
    v_F f(\Lambda)=-L[E^{O}(L,\Lambda) - L\epsilon-\Tilde{b}].
\end{equation}

\begin{figure}
	\centering{}\includegraphics{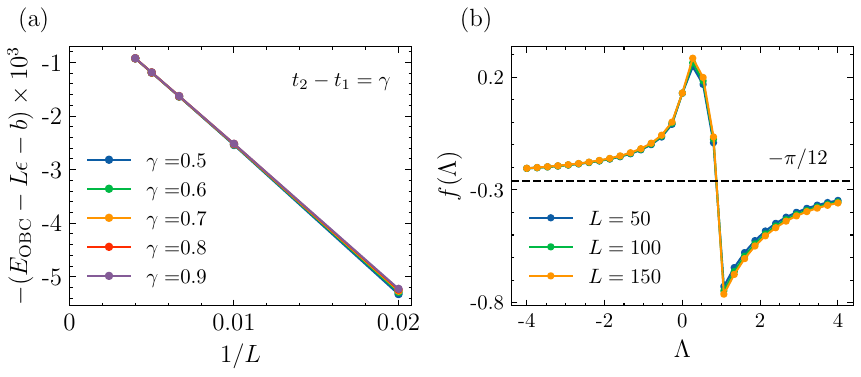}
	
	\caption{\label{Fig2}(a) Finite size scaling of boundary excitation energies at the topological phase transition points $t_2-t_1=\gamma$ with various magnitude of non-Hermicity $\gamma$. The data collapses onto a universal function characterized by a consistent slope of $-\pi/12$, revealing the presence of a negative central charge $c=-2$. (b) Numerical calculation of scaling function $f(\Lambda)$ across different system sizes $L$, with the non-Hermitian scale $\Lambda=L(t_2-t_1)/t_2$ and $t_2-t_1=\gamma$. $f(\Lambda=0)=\pi/24$, consistent with the unitary CFT characterized by a central charge $c=1$. As $|\Lambda|$ increases, $f(\Lambda)$ asymptotically approaches $-\pi/12$, indicating the negative central charge $c=-2$ associated with the non-unitary CFT. The intriguing behavior of the scaling function on different sides of the transition sheds light on the characteristic features of non-Hermitian topological criticality. The data collapse across various system sizes and agrees with the analytic scaling function derived from continuous models.}
\end{figure}

\subsection{Numerical results }
\subsubsection{Negative central charge}
We numerically show that the finite-size scaling of the ground state energies in Eq.\ref{eq:critical-scaling} still holds at the transition points in the non-Hermitian system, with a negative central charge $c=-2$ obtained under the OBC. The Fig.~\ref{Fig2}(a) shows the finite size scaling of the universal part of the boundary excitation for different $\gamma$ at the topological phase transition points $t_2-t_1=\gamma$. As previously discussed, the extraction of universal boundary excitation data necessitates the subtraction of the model-specific contribution $E^{O}-L\epsilon -b$, with the energies measured in units of the Fermi velocity $v_F=\sqrt{t_1 t_2}$. The data for the boundary excitation collapses into the same line as a linear function of $1/L$, with a slope of $\pi/12$. This result indicates the central charge at the topological quantum criticality $t_2-t_1=\gamma$ is $c=-2$.

Upon establishing the exhibition of a negative central charge $c=-2$ at non-Hermitian topological criticality $t_2-t_1=\gamma$, it is imperative to explore the quantum criticality on the topologically trivial side, where $t_2-t_1=-\gamma$. Moreover, it is of paramount importance to delineate the system's evolution from a unitary conformal field theory (CFT) to a non-unitary variant upon the activation of non-Hermitian parameters.

\subsubsection{Universal scaling function}
We further calculate the universal scaling function $f(\Lambda)$ in Eq.~\ref{eq:universalFunc}, which characterizes the flow from the unitary CFT to the non-unitary CFT, with $\Lambda=L(t_2-t_1)/t_2$ and $\gamma=t_2-t_1$. Fig.~\ref{Fig2}(b) shows the scaling function $f(\Lambda)$ calculated in different system sizes under OBC. When  $\Lambda=0$, the system reduces to Hermitian quantum criticality $t_2-t_1=0$ . The scaling function $f(\Lambda=0)=\pi/24$, which is consistent with the central charge $c=1$ for the Hermitian SSH model. As $|\Lambda|$ becomes large,  $f(\Lambda)$ asymptotically approaches $-\pi/12$, showing that the CFT is labelled by the negative central charge $c=-2$ in the non-Hermitian regime, applicable to both the topologically non-trivial and trivial sides of the non-Hermitian critical points.

Notably, the scaling function $f(\Lambda)$ exhibits distinct behaviors depending on the side of the transition, as the $|\Lambda|$ increases from zero to finite value. Specifically: For the topological trivial side critical pints $\Lambda<0$, $f(\Lambda)$ continnuously evolves from $\pi/24$ to $-\pi/12$ in a monototic manner, as $|\Lambda|$ varies from zero to sufficiently large values. This evolution signifies the transition from a unitary conformal field theory (CFT) to a non-unitary one. For the topological non-trivial side $\Lambda>0$, on the other hand, $f(\Lambda)$ displays a non-monototic behavior as $\Lambda$ increases from zero. Initially, it rises, then rapidly falls, and subsequently rises again, revealing a universal singular dip at $\Lambda=1$ [Fig.~\ref{Fig2}(b)].

The singularity observed in $f(\Lambda)$ shows the characteristic finite-size behavior of the non-Hermitian topological transition. Recall that at the topological non-trivial critical point $t_2-t_1=\gamma$, a pair of edge mode with purely imaginary energies $E=\pm i \gamma$ emerges under OBC. The localization length $\xi$ of these edge states is given by $\xi=\sqrt{(t_2/\gamma)^2-(t_2/\gamma)}$, and for $t_2\gg\gamma$, $\xi\sim t_2/\gamma$. Consequently, $\Lambda\sim L/\xi$. The singularity in the scaling function signals the appearance of the non-Hermitian topological edge mode within the finite-size system. For $0<\Lambda<1$, the edge modes are scattered by each other and merge into the bulk. For $\Lambda>1$, the edge modes persist and contribute to the scaling function under OBC,  ultimately saturating $f(\Lambda)$ to $-\pi/12$ when the system size is much larger than the localization length. This result unveils a profound connection between the non-Hermitian topological edge mode and unique quantization condition inherent in the non-unitary CFT, which we explore in detail in the subsequent discussion.

\section{Low-energy theory}
In this section, we derive the analytic expressions of finite-size scaling functions $f(\Lambda)$ in Eq.~\ref{eq:universalFunc}. Our analysis aims to deepen our understanding of the numerical results obtained above. We begin by deriving the quantization conditions from the continuous low-energy Hamiltonian. Subsequently, we explore these quantization conditions under various limits, revealing both the intriguing negative charge change $c=-2$ and the universal non-Hermitian flow of the scaling function.

\subsection{Quantization condition}
\begin{figure}
		\includegraphics[scale=0.8]{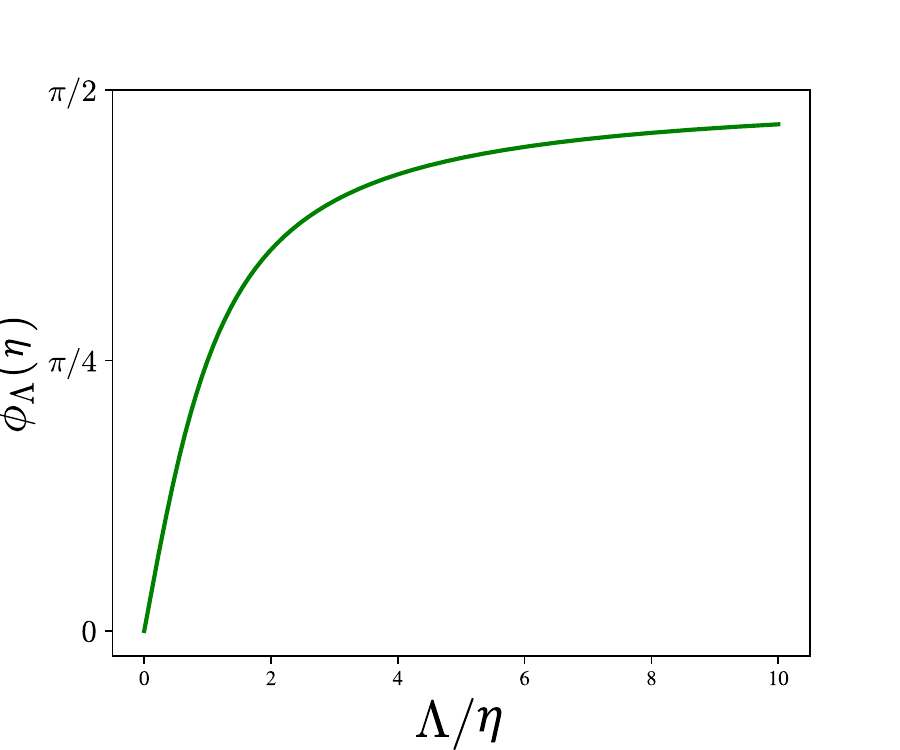}
		\centering
		\caption{(Color online) $\phi_{\Lambda}(\eta)$ as a function of $\Lambda/\eta$, with $\Lambda=L(t_2-t_1)/t_2$, $t_2-t_1=\gamma$ and $\eta=kL$. The phase determines {the quantization condition and further determines} the finite-size spectrum of gapless modes in the flow from the Hermitian to the non-Hermitian fixed point.
		} 
		\label{phase}  
\end{figure}

{We first derive the quantization condition for the critical point, which gives the legitimate momenta contributing to the ground state energy.} The lattice model closes its gap at $k=0$ when $\gamma':= t_2-t_1=\gamma$. The low-energy Hamiltonian {derived} from the expansion around $k=0$ is given by 
\begin{eqnarray}
\label{1}
H(-i\partial_x) = i\gamma\sigma_z + \left(-\gamma' + \frac{t_2}{2}(-i\partial_x)^2 \right)\sigma_x + t_2(-i\partial_x) \sigma_y  .
\end{eqnarray}
{Two key parameters characterize the low-energy theory: $\gamma$ and $t_2$.}
When $|\gamma'| = \gamma$, the system becomes gapless with the Fermi velocity $v_F = \sqrt{-\gamma' t_2 + t_2^2}$. The parameter $\gamma$ is the origin of the non-Hermitian effect. In the Hermitian case ($\gamma = 0$), the $k^2$ term can be neglected, but with nonzero $\gamma$, the quadratic term $k^2$ contributes to the Fermi velocity. The continuous Hamiltonian $H(-i\partial_x)$ operates on the interval $(0,L)$ with OBC $\psi_A(0) = \psi_B(L) = 0$. The unnormalized right-vector is given by:
    \begin{eqnarray}
        \psi_s(k) = (i\gamma + sv_F |k|, -\gamma' + \frac{t_2}{2} k^2 + it_2 k), \quad s=\pm.
    \end{eqnarray}
{This is also valid for the edge state. Since for the edge state with purely imaginary energies $E = iv_F \kappa$, the wavevector can be chosen to be $k = i\kappa$.} The wave function with energy $+v_F|k|$ can be expressed as the linear combination below:
\begin{eqnarray}
    \Psi_k(x) = a\psi_+(k)e^{ikx} + b\psi_+(-k)e^{-ikx} 
\end{eqnarray}
where $a$ and $b$ are arbitrary complex numbers. Now we impose the open boundary condition,  we obtain the following equations:
\begin{eqnarray}
    &\text{left boundary:}& \quad a(i\gamma + v_F |k|) + b(i\gamma + v_F |k|) = 0, \nonumber \\ 
    &\text{right boundary:}& \quad a(-\gamma' + \frac{t_2}{2}k^2 + it_2k)e^{ikL} + b(-\gamma' + \frac{t_2}{2}k^2 - it_2k)e^{-ikL} = 0. \nonumber
\end{eqnarray}
Combining these equations yields the quantization of momentum:
\begin{eqnarray}
    \tan kL = -\frac{k}{-\gamma'/t_2 + k^2/2} \label{6}.
\end{eqnarray}
Notice that $\gamma'/t_2$ is the only parameter controlling the quantization condition. We denote $\eta = kL$ and then Eq.~\ref{6} is rewritten as:
\begin{eqnarray}
    \tan \eta = \frac{\eta}{L\gamma'/t_2 - \eta^2/2L} \label{7}
\end{eqnarray}
Now we consider such a scaling situation: $\Lambda = L\gamma'/t_2$ remains constant while $L\rightarrow \infty$.
Then the quantization condition becomes
\begin{eqnarray}
    \label{qc}
    \tan \eta = \frac{\eta}{\Lambda}.
\end{eqnarray}
This quantization condition determines the low-energy spectrum in the non-Hermitian flow. Below, we consider two asymptotic limits:
\begin{itemize}
    \item {  $\Lambda = 0$: Hermitian case.} The quantization condition simply gives $\eta = (2n+1)\pi/2$, where $n \in \mathbb{Z}$. This is the standard Ising CFT quantization\cite{fran12conformal}.  
    \item {  $\Lambda = \pm \infty$: non-Hermitian case.} 
{The quantization condition leads to the following exotic results:}
    \begin{itemize}
        \item $\Lambda > 0$: $\eta = n\pi$, where $n \in \mathbb{Z}^+$, and $\eta = i\Lambda$.
        \item $\Lambda < 0$: $\eta = n\pi$, where $n \in \mathbb{Z}^+$.
		
		{This intriguing quantization defines the CFT spectrum in PT-symmetric non-Hermitian criticality, ultimately yielding a central charge of $-2$.}
    \end{itemize} 
    
    \item { Finite $\Lambda$: the non-Hermitian flow from unitary CFT to non-unitary CFT.} 
    One can rewrite the quantization condition as:
    \begin{eqnarray}
        \cos\left[\eta + \phi_\Lambda(\eta)\right] = 0
    \end{eqnarray}
    where the phase $\phi$ is defined by $\tan \phi=\Lambda/\eta$.
    {Notably, $\phi_{\Lambda=0}(\eta) = 0$ and $\phi_{\Lambda=\infty}(\eta) = \pi/2$, corresponding to central charges $c=1$ and $c=-2$, respectively.} And when $\Lambda/\eta \gg 1$, the phase behaves as: $\frac{d}{d(\Lambda/\eta)} \phi_\Lambda(\eta) \simeq \frac{\eta^2}{\Lambda^2}$. The phase $\phi_\Lambda(\eta)$ is plotted in Fig.~\ref{phase}. 
 
\end{itemize}

For the boundary mode, {the momentum becomes purely imaginary $k=i \kappa$,} and the quantization condition becomes
    \begin{eqnarray}
        \tanh \kappa L = \frac{i\kappa}{\gamma/t_2 - \kappa^2/2},
    \end{eqnarray}
which gives the localization length of boundary modes in the finite-size system.

\subsection{Negative central charge}
{In this subsection, we prove the negative central charge of the low-energy model in the non-Hermitian regime by tracing the ground state energy with $|\Lambda|\rightarrow\infty$ under the OBC. }
Notice that we focus on the quantum criticalities and thus the dispersion is always linear. The ground state {measured in units of the Fermi velocity} is expressed as the contour integral
\begin{eqnarray}
    E_{\text{G.S.}}/v_F = -\oint_C \frac{dk}{2\pi i} \, k \, \partial_k \ln f(k), 
\end{eqnarray}
where $f(k)= \sin kL$ is the quantization function that gives the legitimate momenta. The contour only encircles the positive axis so that one does not need to bother with double-counting.

Now let us focus on the upper plane contour and write $\sin kL = (e^{ikL} - e^{-ikL})/2i$, 
{yielding} the contribution:
\begin{eqnarray}
    E_+ \equiv -v_F \int_{i\delta}^{i\delta+\infty} \frac{dk}{2\pi i} k \, \partial_k \ln \frac{e^{ikL} - e^{-ikL}}{2i},
\end{eqnarray}
{where the extraction of $e^{-ikL}$ from the logarithm imparts the bulk energy. Subsequently, we examine the remaining component:}
\begin{eqnarray}
    J_+ \equiv -v_F \int_{i\delta}^{i\delta+\pi} \frac{dk}{2\pi i} k \, \partial_k \ln  (e^{2ikL} - 1).
\end{eqnarray}
Here we also omit the factor $2i$ from the logarithm as it does not contribute to the result. By contour deformation, the integral 
{transmutes} to:
\begin{eqnarray}
    J_+ \equiv \frac{v_F}{L} \int_{0}^{\infty} \frac{dz}{2\pi} z \partial_z \ln (e^{-2z} - 1) = -\frac{v_F}{L} \int_{0}^{\infty} \frac{dz}{\pi} \frac{z}{1 - e^{2z}}=\frac{v_F}{L} \frac{\pi}{24},
\end{eqnarray}
where we have treated $L\pi$ as infinity. 
An exactly same contribution arises from the lower half-plane.
Thus we reach the conclusion below:
\begin{eqnarray}
    E_{\text{G.S.}} = N\epsilon + b + \frac{\pi}{12} \frac{v_F}{L} + O(L^{-2}).
\end{eqnarray}
Comparing to the standard CFT result:
\begin{eqnarray}
    E_{\text{G.S.}} = N\epsilon + b - \frac{\pi}{24} \frac{v_F c}{L} + O(L^{-2}),
\end{eqnarray}
we conclude that the CFT described by Eq.~\ref{1} carries the negative central charge $c = -2$.
On the contrary, under PBC, the quantization of $k$ is given by $2\pi/L$, and the central charge computed in the PBC is $c=1$. Therefore, the realization of non-unitary CFT in the lattice model is very sensitive to boundary conditions.

\subsection{Universal scaling function }
We further derive the universal scaling function associated with the non-Hermitian flow by evaluating the ground state energy for a finite $\Lambda$ under OBC. The ground state energy in the unit of the Fermi velocity $v_F$ is expressed as
\begin{eqnarray}
E_{G.S.}/v_F=-{(2\pi i)}^{-1}\oint_C   dk \,   k \, \partial_k \ln f(k),
\end{eqnarray}
where the quantization function is {defined} by $
f(k)=    \cos (kL + \phi). $ Distinct quantization conditions are obtained for $\Lambda>0$ and $\Lambda<0$, necessitating separate treatments.
\begin{itemize}
    \item Firstly we focus on the case $\Lambda<0$. In this case, the contour encompasses positive $x$-axis.  Now we focus on the upper plane contour and find its contribution
\begin{eqnarray}
E_+/v_F=-(2\pi i)^{-1} \int_{i\delta}^{i\delta+\pi}  {dk}  \,   k\,  \partial_k \, \ln  \Big( k  \cos kL -  \gamma  \sin kL\Big).
\end{eqnarray}
We first extract out $e^{-ikL}$ out the logarithm since this part contributes to the bulk energy and then consider the residual term
\begin{eqnarray}
E_+/v_F= \text{bulk part} -(2\pi i)^{-1} \int_{i\delta}^{i\delta+\pi}  {dk}  \,   k\,  \partial_k \, \ln  \Big[ k (e^{2ikL}+1) -  \gamma  (e^{2ikL}-1)/i\Big].
\end{eqnarray}
After deforming the contour to the imaginary axis, the integral becomes
\begin{eqnarray}
E_+/v_F= \text{bulk part} +(2\pi L )^{-1} \int_{0}^{+\pi L}  {dx}  \,   x\,  \partial_x \, \ln  \Big[ x  (e^{-2x}+1) +  \Lambda (e^{-2x}-1)\Big].
\end{eqnarray}
One can perform derivatives and find
\begin{eqnarray}
E_+/v_F= \text{bulk part} +(2\pi L )^{-1} \int_{0}^{ \pi L}  {dx}  \,   x\,  \,    \frac{ 1-2x-2 \Lambda + e^{2x} }{(x+ \Lambda )   + (x- \Lambda )e^{2x}  }.
\end{eqnarray}
We then extract the boundary energy out of the integral. 
Representing the integral in terms of the $k$ variable temporarily, we have
\begin{eqnarray}
\int_{0}^{ \pi L}  \frac{dx}{2\pi L }  \,   x\,  \,    \frac{ 1-2x-2 \Lambda + e^{2x} }{(x+ \Lambda )   + (x- \Lambda )e^{2x}  }=  \int_{0}^{ \pi } \frac{dk}{2\pi} \,   k\,  \,  \frac{ (1-2k-2 \gamma) e^{-2Lk}+ 1 }{(k+\gamma )e^{-2Lk}   + (k- \gamma )   }.
\end{eqnarray}
Polynomial decomposition is applied as follows:
\begin{eqnarray}
  \frac{ (1-2k-2 \gamma)  + x }{k+\gamma   + (k- \gamma )x  }=\frac{1}{k-\gamma}+\frac{-(k+\gamma)/(k- \gamma )+(1-2k-2 \gamma)  }{(k+\gamma+ (k- \gamma )x      )}.
\end{eqnarray}
The first term contributes to the boundary term, while the remaining part accounts for the finite-size correction that we focus on. Consequently, we derive the scaling function {for $\Lambda<0$ as}
\begin{eqnarray}
\label{eq:f1}
f(\Lambda)\equiv ( \pi   )^{-1} \int_{0}^{ \infty}  {dx}  \,     \,  x  \frac{-(x+\Lambda)/(x- \Lambda )+(1-2x-2 \Lambda)  }{ x+\Lambda+ (x- \Lambda )e^{2x}      },\quad \Lambda<0.
\end{eqnarray}

 \item Now we consider positive $\Lambda\geq 0$. This case needs to be carefully treated, since the quantization function allows the imaginary solution when $\Lambda>1$. 
{The imaginary momenta satisfy the quantization condition}
\begin{eqnarray}
\tanh \kappa=\frac{\kappa}{\Lambda}.
\end{eqnarray}
When $\Lambda\gg 1$, $\kappa$ approximates to $\Lambda$. 
The imaginary solution results in the imaginary energy. However, since we only account for the real part of the energy and exclude imaginary energy corrections, we maintain the same contour and evaluate the integral as
\begin{eqnarray}
E_+/v_F=-(2\pi i)^{-1} \int_{i\delta}^{i\delta+\pi}  {dk}  \,   k\,  \partial_k \, \ln  \Big( k  \cos kL -  \gamma  \sin kL\Big).
\end{eqnarray}
The procedures are analogous, but the integral along imaginary axis will be taken to be principal value of integral. That is
\begin{eqnarray}
E_+/v_F= \text{bulk part} +(2\pi L )^{-1} \text{P} \int_{0}^{ \pi L}  {dx}  \,   x\,  \,    \frac{ 1-2x-2 \Lambda + e^{2x} }{(x+ \Lambda )   + (x- \Lambda )e^{2x} }.
\end{eqnarray}
Here \text{P} indicates the principal integral. 
Through similar calculations, we obtain {the scaling function for $\Lambda>0$ as}
\begin{eqnarray}
\label{eq:f2}
f(\Lambda)\equiv   \text{P}\, \frac{1}{\pi}\int_{0}^{ \infty}  {dx}  \,     \,  x  \frac{-(x+\Lambda)/(x- \Lambda )+(1-2x-2 \Lambda)  }{ x+\Lambda+ (x- \Lambda )e^{2x}      }, \quad \Lambda>0.
\end{eqnarray}

\end{itemize}

\begin{figure}
		\includegraphics[scale=0.9]{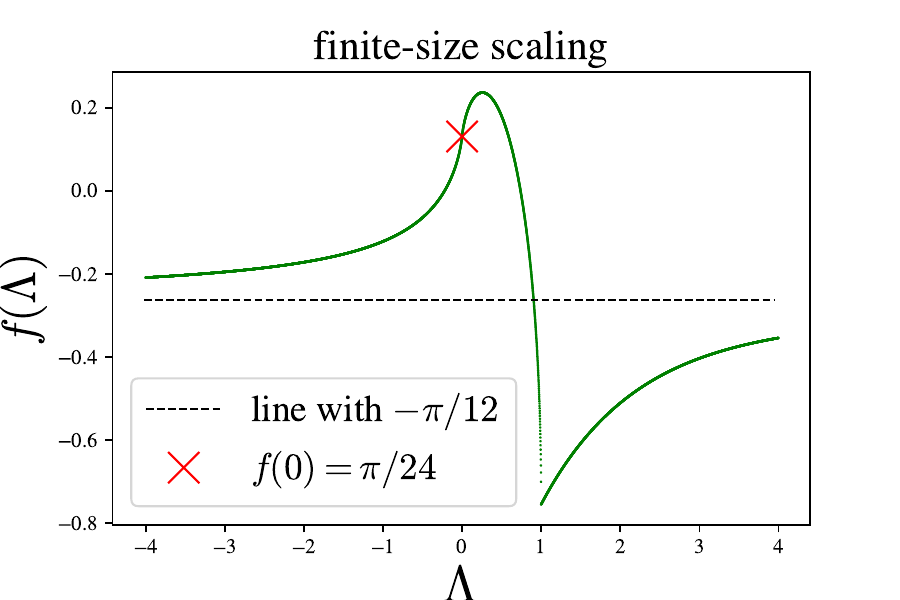}
		\centering
		\caption{(Color online) Analytic scaling function $f(\Lambda)$ versus $\Lambda$. {A notable dip at $\Lambda=1$ is observed, attributed to the qualitative shift in the quantization condition for momenta, associated with universal singularity at $\Lambda=1$. For $\Lambda<1$ quantization condition yields real solutions, whereas imaginary solution emerges for $\Lambda>1$, giving rise to the rise-dip-rise behavior within the topologically non-trivial side of non-Hermitian criticality.} 
		} 
		\label{plot}  
\end{figure}

{The universal scaling functions, combined by Eq.~\ref{eq:f1} and Eq.~\ref{eq:f2}, are depicted in Fig.~\ref{plot}. Notice that when numerically plotting the scaling functions, one needs to keep the integral above avoiding two poles: $x=\Lambda$ and $\tanh x=x/\Lambda$. The analytic and numerical results are in good agreement. Fig.~\ref{plot} illustrates the analytic non-Hermitian flow from the unitary CFT to the non-unitary CFT, consistent with the numerical findings. The distinct behaviors of the scaling functions on the topologically trivial and non-trivial sides highlight the qualitative shift in the quantization condition for momenta. This shift explains the universal singularity at the dip $\Lambda=1$ within the rise-dip-rise pattern on the topologically non-trivial side. The universal singularity originates from the emergence of imaginary solution to the quantization condition when $\Lambda>1$, signifying the onset of dissipative boundary modes with imaginary momentum. Conversely, when $\Lambda<1$, all solutions of the quantization condition are real, indicating that all eigenstates are bulk modes with real momenta, despite the non-Hermitian nature of the system. Therefore, the universal dip at $\Lambda=1$ epitomizes the {\it characteristic} nature of non-Hermitian topological transition, marking the inception of dissipative topological boundary modes in the spectrum and indicating the system's flow towards the non-unitary CFT characterized by $c=-2$.}

\section{Conclusions and discussions} 
We have studied the $\mathcal{PT}$-symmetric quantum criticality via both lattice model and continuous low-energy theory. We analytically derived a universal scaling function capturing the non-Hermitian flow from unitary to non-unitary conformal field theories (CFTs). Our analytical results reveal the universality of this scaling function across a wide range of PT-symmetric non-Hermitian systems with linear dispersion, independent of the specific microscopic details of the model. Notably, this universal scaling function not only displays the non-Hermitian corrections to spectrum, but also exhibits exotic rise-dip-rise pattern attributed to singularity on the topological transition side. The distinct behavior of the scaling function at different critical points that separate the $\mathcal{PT}$-broken and $\mathcal{PT}$ symmetric phases, demonstrates its sensitivity to the intrinsic topology of the non-Hermitian quantum criticality. This sensitivity enables us to distinguish between transitions from $\mathcal{PT}$-broken to $\mathcal{PT}$-trivial phases and transitions from $\mathcal{PT}$-broken to $\mathcal{PT}$-topological phases.

Importantly, our findings, in conjunction with the preceding study by Chang et al.~\cite{chang2020}, highlight the sensitivity of the emergence of non-unitary CFT to boundary conditions. In our research, focusing on the ground state energy, we observe the manifestation of a negative central charge $c=-2$ under open boundary condition (OBC), whereas the central charge reverts to $c=1$ under periodic boundary condition (PBC). Conversely, in the study conducted by Chang et al.~\cite{chang2020}, which examines entanglement entropy, the negative central charge transitions to a positive value, $c=1$, under OBC, while it assumes a negative value, $c=-2$, under PBC. This elucidates that the non-unitary information of a quantum system with OBC is encoded within the conformal spectrum, whereas in the context of entanglement entropy, which deals with the quantum states, the non-unitary information is embedded in the unnormalizable quantum state, only emerging for periodic boundary condition (PBC).

\section*{Acknowledgements}
We thank Xiong-Jun Liu and Yuhan Liu for insightful discussions.

\paragraph{Funding information}
X.-C. Zhou was supported by National Key Research and Development Program of China (2021YFA1400900), the National Natural Science Foundation of China (Grants No. 11825401 and No. 12261160368), and the Innovation Program for Quantum Science and Technology (Grant No. 2021ZD0302000). K. Wang was supported, in part, by the Kadanoff Center for Theoretical Physics.
	\nolinenumbers
	 \bibliography{nonunitaryCFT}
	
\end{document}